\documentclass[
 aip,
 amsmath,amssymb,
 preprint,%
]{revtex4-2}

\usepackage{graphicx}
\usepackage{dcolumn}
\usepackage{bm}
\usepackage[paperwidth=210mm,paperheight=297mm,centering,hmargin=1cm,vmargin=2.5cm]{geometry} 
\usepackage[utf8]{inputenc}
\usepackage[T1]{fontenc}
\usepackage{mathptmx}

\usepackage{amsmath}
\usepackage{mathtools, cuted}
\usepackage{lineno,hyperref}
\usepackage{xcolor}
\modulolinenumbers[5]

\usepackage{blindtext}

\newcommand{\ket}[1]{\left|#1\right\rangle}
\newcommand{\bra}[1]{\left\langle#1\right|}

\begin{document}
\title{Description of interface between semiconductor electrostatic qubit and Josephson junction in tight binding model }

\author{Krzysztof Pomorski$^{1,2,3,4}$}   \email[Corresponding author:]{kdvpomorski@gmail.com}
 \affiliation{
  1:University College Dublin-School of Computer Science, Ireland, \\
  2:University College Dublin-School of Electrical and Electronic Engineering, Ireland, \\
  3: Technical University of Lodz, Department of Microelectronics and Computer Science, Lodz, Poland \\
  4: Quantum Hardware Systems $(www.quantuhardwaresystems.com)$
}

\begin{abstract}
The interface between superconducting Josephson junction and semiconductor position-based qubit implemented in coupled semiconductor q-dots is described such that it can be the base for electrostatic interface between superconducting and semiconductor quantum computer. Modification of Andreev Bound State in Josephson junction by the presence of semiconductor qubit in its proximity and electrostatic interaction with superconducting qubit is spotted by the minimalist tight-binding model. The obtained results allow in creating interface between semiconductor quantum computer and superconducting quantum computer. They open the perspective of construction of QISKIT like software that will describe both types of quantum computers as well as their interface.
\end{abstract}

\maketitle

\begin{figure}
\centering
\label{fig:central}
\includegraphics[scale=0.5]{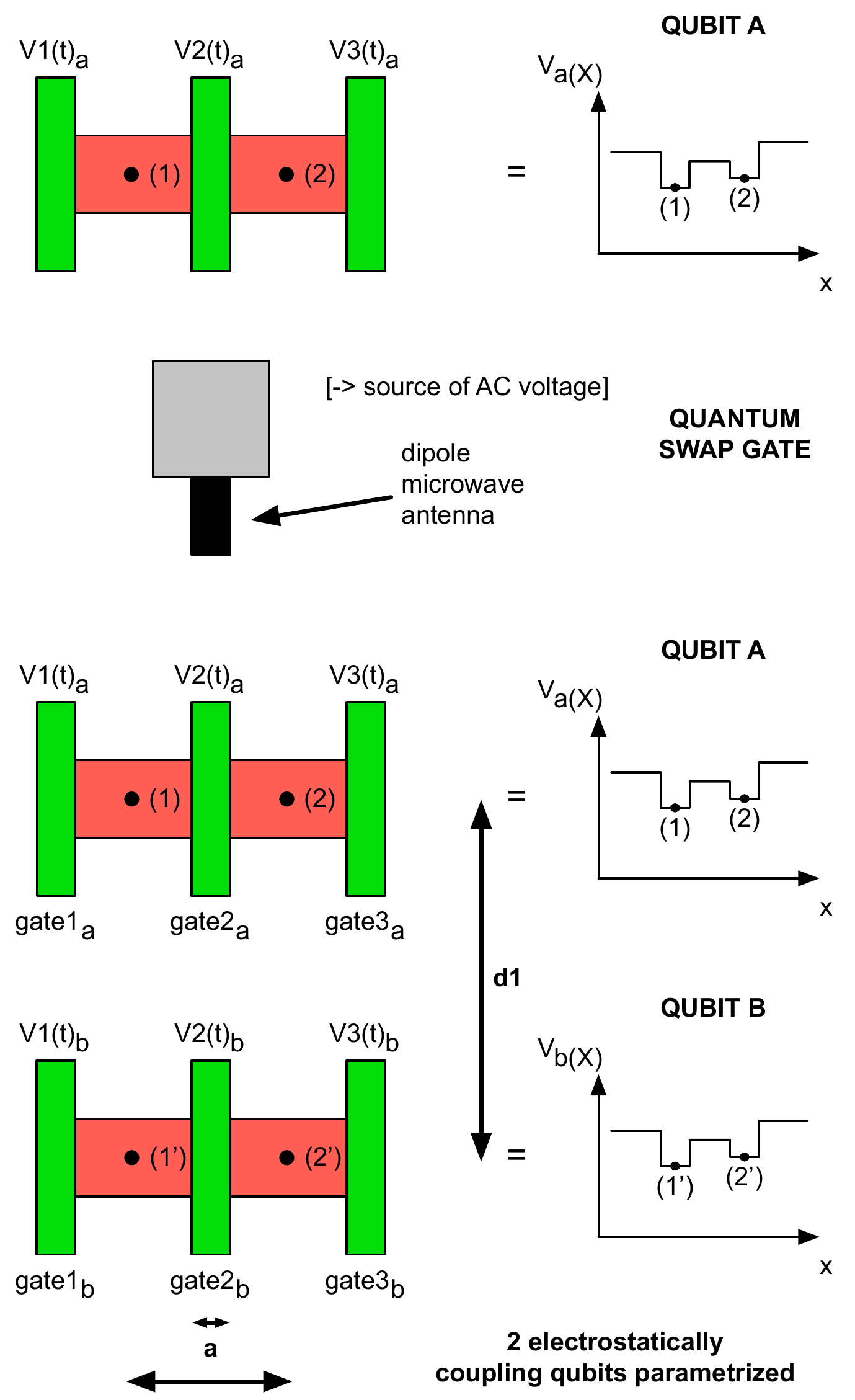} 
\caption{[Left]: Electrostatic position-based qubit implemented in CMOS technology \cite{Pomorski_spie}. [Upper]: Simplistic representation by particle localized in two regions of space denoted by nodes (1) and (2); [Lower]: Case of two electrostatically interacting qubits implementing quantum swap gate. Quantum dynamics are parameterized by presence of electrons at nodes 1, 2, 1' and 2'.}
\end{figure}
\section{Description of position based-qubit in tight-binding model}
We refer to the physical situation from Fig.1 and we consider position based-qubit in tight-binding model \cite{SEL} and its the Hamiltonian of this system is given as
\begin{eqnarray}
\label{simplematrix}
\hat{H}(t)=
\begin{pmatrix}
E_{p1}(t) & t_{s12}(t) \\
t_{s12}^{\dag}(t) & E_{p2}(t)
\end{pmatrix}_{[x=(x_1,x_2)]}= 
(E_1(t)\ket{E_1}_t \bra{E_1}_t+E_2(t)\ket{E_2}\bra{E_2})_{[E=(E_1,E_2)]}.
\end{eqnarray}
The Hamiltonian $\hat{H}(t)$ eigenenergies $E_1(t)$ and $E_2(t)$ with $E_2(t)>E_1(t)$ are given as
\begin{eqnarray}
E_1(t)= \left(-\sqrt{\frac{(E_{p1}(t)-E_{p2}(t))^2}{4}+|t_{s12}(t)|^2}+\frac{E_{p1}(t)+E_{p2}(t)}{2}\right), \nonumber \\
E_2(t)= \left(+\sqrt{\frac{(E_{p1}(t)-E_{p2}(t))^2}{4}+|t_{s12}(t)|^2}+\frac{E_{p1}(t)+E_{p2}(t)}{2}\right),
\end{eqnarray}
and energy eigenstates $\ket{E_1(t)}$ and $\ket{E_2(t)}$ have the following form
\begin{eqnarray}
\ket{E_1,t}=
\begin{pmatrix}
\frac{(E_{p2}(t)-E_{p1}(t))+\sqrt{\frac{(E_{p2}(t)-E_{p1}(t))^2}{2}+|t_{s12}(t)|^2}}{-i t_{sr}(t)+t_{si}(t)} \\
-1
\end{pmatrix},  \nonumber \\
\ket{E_2,t}=
\begin{pmatrix}
\frac{-(E_{p2}(t)-E_{p1}(t))+\sqrt{\frac{(E_{p2}(t)-E_{p1}(t))^2}{2}+|t_{s12}(t)|^2}}{t_{sr}(t) - i t_{si}(t)} \\
1
\end{pmatrix}.
\end{eqnarray}
This Hamiltonian gives a description of two coupled quantum wells as depicted in Fig.1.
In such situation we have real-valued functions $E_{p1}(t)$, $E_{p2}(t)$ and complex-valued functions $t_{s12}(t)=t_s(t)=t_{sr}(t)+i t_{si}(t)$ and $t_{s21}(t)=t_{s12}^{*}(t)$, what is equivalent to the knowledge of four real valued time-dependent continuous or discontinues functions $E_{p1}(t)$, $E_{p1}(2)$ , $t_{sr}(t)$ and $t_{si}(t)$. The quantum state is a superposition of state localized at node 1 and 2 and therefore is given as
\begin{equation}
\ket{\psi}_{[x]}=\alpha(t)\ket{1,0}_x+\beta(t)\ket{0,1}_x=
\alpha(t)
\begin{pmatrix}
1 \\
0 \\
\end{pmatrix}
+
\beta(t)
\begin{pmatrix}
0 \\
1 \\
\end{pmatrix} ,
\end{equation}
where $|\alpha(t)|^2$ ($|\beta(t)|^2$) is probability of finding particle at node 1(2) respectively, which brings $|\alpha(t)|^2+|\beta(t)|^2=1$ and obviously $\bra{1,0}_x\ket{|1,0}_x=1=\bra{0,1}_x\ket{|0,1}_x$ and $\bra{1,0}_x\ket{|0,1}_x=0=\bra{0,1}_x\ket{|1,0}_x$. In Schr\"odinger formalism, states $\ket{1,0}_x$ and $\ket{0,1}_x$ are Wannier functions that are parametrized by position $x$. We work in tight-binding approximation and quantum state evolution with time as given by
\begin{equation}
 i \hbar \frac{d}{dt}\ket{\psi(t)}=\hat{H}(t)\ket{\psi(t)}=E(t)\ket{\psi(t)}.
\end{equation}
The last equation has an analytic solution
\begin{equation}
\ket{\psi(t)}=e^{\frac{1}{i \hbar}\int_{t_0}^{t}\hat{H}(t_1)dt_1}\ket{\psi(t_0)}=e^{\frac{1}{i \hbar}\int_{t_0}^{t}\hat{H}(t_1)dt_1}
\begin{pmatrix}
\alpha(0) \\
\beta(0) \\
\end{pmatrix}
\end{equation}
and in quantum density matrix theory we obtain
\begin{eqnarray}
\hat{\rho}(t)=\hat{\rho}^{\dag}(t)=\ket{\psi(t)}\bra{\psi(t)}= 
\hat{U}(t,t_0)\hat{\rho}(t_0)\hat{U}(t,t_0)^{-1} 
=e^{\frac{1}{i \hbar}\int_{t_0}^{t}\hat{H}(t_1)dt_1}(\ket{\psi(t_0)}\bra{\psi(t_0)})e^{-\frac{1}{i \hbar}\int_{t_0}^{t}\hat{H}(t_1)dt_1}= \nonumber \\
=e^{\frac{1}{i \hbar}\int_{t_0}^{t}\hat{H}(t_1)dt_1}
\bigg(
\begin{pmatrix}
\alpha(0) \\
\beta(0) \\
\end{pmatrix} 
\begin{pmatrix}
\alpha^{*}(0) & \beta^{*}(0) \\
\end{pmatrix}
\bigg)e^{-\frac{\int_{t_0}^{t}\hat{H}(t_1)dt_1}{i \hbar}} 
=\hat{U}(t,t_0)  
\begin{pmatrix}
|\alpha(0)|^2 & \alpha(0)\beta^{*}(0)  \\
\beta(0)\alpha(0)^{*} &  |\beta(0)|^2 \\
\end{pmatrix} \hat{U}(t,t_0)^{\dag}.
\end{eqnarray}

\section{Electrostatic interaction of Josephson junction qubit with semiconductor electrostatic qubit}
The state of Josephson junction is well described by Bogoliubov-de Gennes (BdGe) equation \cite{PSSB2012} pointing the correlation between electron and holes as
\begin{equation}
\begin{pmatrix}
H_0 & \Delta(x) \\
\Delta(x)^{*} & -H_0^{\dag} \\
\end{pmatrix}
\begin{pmatrix}
u_n(x) \\
v_n(x) \\
\end{pmatrix}
=E_n
\begin{pmatrix}
u_n(x) \\
v_n(x) \\
\end{pmatrix} ,
\end{equation}
where $H_0=-\frac{\hbar^2}{2m}\frac{d^2}{dx^2}$ is free electron Hamiltonian with self-consistency relation  $\Delta(x)=\sum_n(1-2f(E_n))u_n(x)v_n(x)^{*}$, where $\Delta(x)$ is the superconducting order parameter and $f(E_n)=\frac{1}{1+e^{-\frac{E_n}{k_bT}}}$ is Fermi-Dirac distribution function and $u_n(x)$ and $v_n(x)$ are electron and hole wavefunctions. In case of bulk superconductor with constant superconducting order parameter we obtain $E_n=\pm \sqrt{|H_0|^2+|\Delta|^2}$.
In later considerations we are going to omit the self-consistency relation assuming the depedence of superconducting order parameter as step-like function. It shall be underlined that BdGe equation is mean field
equation that is dervied basing on BCS theory of superconductivity. It it thus naturally valid for the case of many particles.
Semiconductor single electron line with 2 nodes can be regarded as electrostatic position dependent qubit and can be described by
$H_{semi}=t_{s1,2}\ket{1}\bra{2}+t_{s2,1}\ket{2}\bra{1}+E_{p1}\ket{1}\bra{1}+E_{p2}\ket{2}\bra{2}, $

\begin{figure}
\centering
\label{fig:JJvssemi}
\includegraphics[scale=0.5]{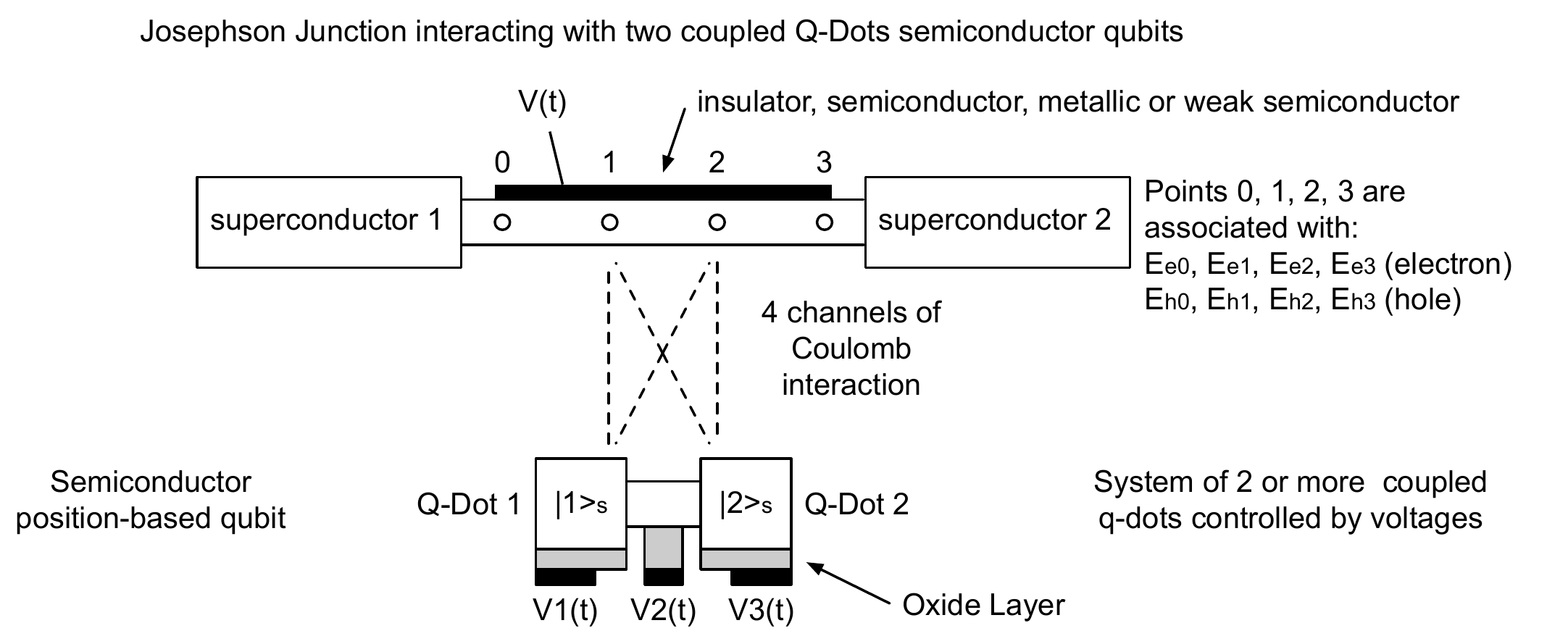} 
\caption{Superconducting Josephson junction interacting with semiconductor position based qubit in minimalistic tight-binding approach, where tight-binding BdGe equation describing Josephson junction is coupled electrostatically to tight-binding model of semiconductor position based qubit.}
\end{figure}

We refer to the physical situation depicted in Fig.2. We can express coupling of 2 systems assuming 4 nodes for electron or hole and 2 nodes for electron confined in semiconductor so we have eigenvector having 16 components ($\ket{0}_e\ket{1}_s$, $\ket{0}_e\ket{2}_s$,$\ket{1}_e\ket{1}_s$, $\ket{1}_e\ket{2}_s$,$\ket{2}_e\ket{1}_s$, $\ket{2}_e\ket{2}_s $, $ \ket{3}_e\ket{1}_s, \ket{2}_e\ket{2}_s $ ), ($\ket{0}_h\ket{1}_s$, $\ket{0}_h\ket{2}_s$,$\ket{1}_h\ket{1}_s$, $\ket{1}_h\ket{2}_s$,$\ket{2}_h\ket{1}_s$, $\ket{2}_h\ket{2}_s $, $ \ket{3}_h\ket{1}_s, \ket{2}_h\ket{2}_s $ )
where s refers to semiconductor qubit whose quantum state is superposition of $
\ket{1}_s$ and $ \ket{2}_s$ and states $\ket{0}_e$, .., $\ket{3}_e$, $\ket{0}_h$, .., $\ket{3}_h$ characterizes the state of electron and hole respectively in ABS [Andreev Bound State when electron moving in normal (non-superconducting) region between superconducors is reflected as hole when it comes into superconducint area and when hole moving in normal region is reflected as electron when it meets superconductor etc .. ] of Josephson junction. This time the quantum state of the system can be written as
\begin{eqnarray}
\ket{\psi,t}= 
\gamma_1(t)\ket{0}_e\ket{1}_s +
\gamma_2(t)\ket{0}_e\ket{2}_s +\gamma_3(t)\ket{1}_e\ket{1}_s+ 
\gamma_4(t)\ket{1}_e\ket{2}_s+\gamma_5(t)\ket{2}_e\ket{1}_s+ 
\gamma_6(t)\ket{2}_e\ket{2}_s+
\gamma_7(t)\ket{2}_e\ket{1}_s \nonumber \\ +\gamma_8(t)\ket{2}_e\ket{2}_s+ 
\gamma_9(t)\ket{0}_h\ket{1}_s+\gamma_{10}(t)\ket{0}_h\ket{2}_s+\gamma_{11}(t)\ket{1}_h\ket{1}_s 
+\gamma_{12}(t)\ket{1}_h\ket{2}_s+\gamma_{13}(t)\ket{2}_h\ket{1}_s+\gamma_{14}(t)\ket{2}_h\ket{2}_s+ 
\nonumber \\ \gamma_{15}(t)\ket{2}_e\ket{1}_s+  
\gamma_{16}(t)\ket{2}_h\ket{2}_s. \nonumber \\
\end{eqnarray}
Normalization condition implies $|\gamma_1(t)|^2+|\gamma_2(t)|^2+..+|\gamma_{16}(t)|^2=1$ at any instance of time t. Such system has 16 eigenenergies. The probability of find electron at node 1 under any presence of electron in semiconductor qubit at node 1 or 2 is obtained by appling projection of $\bra{1}_e\bra{1}_s+\bra{1}_e\bra{2}_s$ so $|\bra{1}_e\bra{1}_s+\bra{1}_e\bra{2}_s \ket{\psi,t}|^2$ is probability of finding electron at node 1 in Josephson junction.
We obtain the following structures of matrices corresponding to $H_0$ part of BdGe equation in the forma as
\begin{eqnarray}
\hat{H}_{0[e]}=
\begin{pmatrix}
E_{p1} + E_{e0}                    & t_s                                             & t_{e(1,0)}                                           & 0                                                           & t_{e(2,0)}                                          & 0                                            & t_{e(3,0)}         & 0                                   \\
t_s^{*}                                    & E_{p2} +  E_{e0}                    & 0                                                          & t_{e(1,0)}                                            & 0                                                         & t_{e(2,0)}                             & 0                       & t_{e(3,0)}                     \\
t_{e(1,0)}^{*}                        & 0                                                & E_{p1} + \frac{q^2}{a} +  E_{e1}   & t_s                                                        & t_{e(2,1)}                                          & 0                                            & t_{e(3,1)}         & 0                                   \\
0                                               & t_{e(1,0)}^{*}                         & t_s^{*}                                                & E_{p2} +  E_{e1} + \frac{q^2}{b}   & 0                                                         & t_{e(2,1)}                             & 0                       & t_{e(3,1)}                    \\
t_{e(2,0)}^{*}                        & 0                                                & t_{e(2,1)}^{*}                                    & 0                                                           & E_{p1} +E_{e2} + \frac{q^2}{b}   & t_s                                         & t_{e(3,2)}        & 0                                   \\
0                                               &  t_{e(2,0)}^{*}                        & 0                                                           & t_{e(2,1)}^{*}                                    & t_s^{*}                                              & E_{p2} + E_{e2}+ \frac{q^2}{a} & 0                       & t_{e(3,2)}                    \\
t_{e(3,0)}^{*}                        & 0                                                &  t_{e(3,1)}^{*}                                   & 0                                                           & t_{e(3,2)}^{*}                                  & 0                                            & E_{p1} + E_{3e}       & t_s                                \\
0                                               & t_{e(3,0)}^{*}                         & 0                                                           &  t_{e(3,1)}^{*}                                   & 0                                                         & t_{e(3,2)}^{*}                     & t_s^{*}            & E_{p2} + E_{3e}           \\
\end{pmatrix} \nonumber \\
\end{eqnarray}
\normalsize
Parameters $E_{p1}$, $E_{p2}$, $t_s$ correspond to semiconductor position based qubit and distance between semiconductor qubit and Josephson junction is given by a and b. Other parameters $E_{e0},E_{e1}, E_{e2},E_{e3}$ , $E_{h0},E_{h1}, E_{h2},E_{h3}$
describes localization energy of electron and hole at nodes 0, 1, 2 and 3 of Josephson junction.
In analogical way we can write
\begin{eqnarray}
\hat{H}_{0[h]}=
\begin{pmatrix}
E_{p1} + E_{h0}                    & t_s                                             & t_{h(1,0)}                                           & 0                                                           & t_{h(2,0)}                                          & 0                                                         & t_{h(3,0)}                 & 0                                   \\
t_s^{*}                                    & E_{p2} +  E_{h0}                    & 0                                                          & t_{h(1,0)}                                            & 0                                                         & t_{h(2,0)}                                          & 0                                & t_{h(3,0)}                     \\
t_{h(1,0)}^{*}                        & 0                                                & E_{p1} - \frac{q^2}{a} +  E_{h1}   & t_s                                                        & t_{h(2,1)}                                          & 0                                                         & t_{h(3,1)}                 & 0                                   \\
0                                               & t_{h(1,0)}^{*}                         & t_s^{*}                                                & E_{p2} +  E_{h1} - \frac{q^2}{b}   & 0                                                         & t_{h(2,1)}                                          & 0                                & t_{h(3,1)}                    \\
t_{h(2,0)}^{*}                        & 0                                                & t_{h(2,1)}^{*}                                    & 0                                                           & E_{p1} +E_{h2} - \frac{q^2}{b}   & t_s                                                      & t_{h(3,2)}                 & 0                                   \\
0                                               &  t_{h(2,0)}^{*}                        & 0                                                           & t_{h(2,1)}^{*}                                    & t_s^{*}                                              & E_{p2} + E_{h2}- \frac{q^2}{a}   & 0                                & t_{h(3,2)}                    \\
t_{h(3,0)}^{*}                        & 0                                                &  t_{h(3,1)}^{*}                                   & 0                                                           & t_{h(3,2)}^{*}                                  & 0                                                         & E_{p1} + E_{3h}       & t_s                                \\
0                                               & t_{h(3,0)}^{*}                         & 0                                                           &  t_{h(3,1)}^{*}                                   & 0                                                         & t_{h(3,2)}^{*}                                  & t_s^{*}                     & E_{p2} + E_{3h}           \\
\end{pmatrix} \nonumber \\
\end{eqnarray}
\normalsize
and two other matrices
$\hat{\Delta}_1=diag(\Delta(0),\Delta(0),\Delta(1),\Delta(1),\Delta(2),\Delta(2),\Delta(3),\Delta(3)),
\hat{\Delta}_2=\hat{\Delta}_1^{\dag}$.
Finally we obtain the following structure of tight-binding Bogoliubov-de Gennes equations including the interaction of semiconductor qubit with Josephson junction described in the minimalistic way in the form
\begin{equation}
\hat{H}_{eff}=
\begin{pmatrix}
\hat{H}_{0[e]} & \hat{\Delta}_1 \\
\hat{\Delta}_2 & \hat{H}_{0[h]}
\end{pmatrix}.
\end{equation}

Similarly as before, having knowledge of quantum state at $t_0$ we can evaluate the state at time $t$ by computing $\exp(\int_{t_0}^{t}\frac{1}{\hbar i}\hat{H}_{ext}(t)dt')\ket{\psi,t_0)}=\ket{\psi,t)}$ which bases on the same method already presented before in Eq.\,(8).
We can also perform the procedure of heating up or cooling down of the quantum state in the way as it was described before or we can regulate the population of pointed energetic level(s).

In most minimalistic tight-binding model of Josephson junction Sc-I-Sc (Superconductor-Insulator-Superconductor) we set $\Delta(1)=\Delta(2)=0$ what corresponds to the simplest form of Andreev Bound State in Tunneling Josephson junction. However in weak-links and in the Field Induced Josephson junctions all diagonal elements are non-zero and $|\Delta|$ has maximum at $\Delta(0)$ and $\Delta(3)$ that can be considered as superconducting state of bulk superconductors. Quite naturally, Field Induced Josephson junction \cite{PSSB2012} can have special profile of dependence of superconducting order parameter $\Delta(x)$ on position x with presence of built-in magnetic fields in area of junction. It will also have special complex-valued hopping constants for electron and hole in area of superconductor that
will incorporate the profile of magnetic field present across Josephson junction.
Specified Hamiltonian describing electrostatic interface between superconducting Josephson junction and semiconductor position-based qubit has the following parameters describing the state of position based semiconductor qubit $E_{p1}$, $E_{p2}$ , $t_s=t_{sr}+i t_{is}$ (4 real valued time dependent functions),
and parameters describing the state of Josephson junction $E_{e0}$, $E_{e1}$, $E_{e2}$,$E_{e3}$, $E_{h0}$,$E_{h1}$, $E_{h2}$,$E_{h3}$ , $\Delta(0)$, $\Delta(1)$, $\Delta(2)$, $\Delta(3)$, $t_{e(1,0)}$, $t_{e(2,1)}$, $t_{e(2,3)}$, $t_{e(3,0)}$, $t_{h(1,0)}$, $t_{h(2,1)}$, $t_{h(2,3)}$, $t_{h(3,0)}$  as well as geometrical parameters describing electrostatic interaction between semiconductor JJ and semiconductor qubit by a and b. It is worth mentioning that electrostatic interaction taken into account is only between nodes 1-1s, 1-2s,2-1s,2-2s what means 4 channels for Coulomb interaction and simplifies the model greatly so one can find analytical solutions as well. The assumption with four channels of electrostatic interaction is physically justifiable if one assumes that $\Delta(0) \neq 0, \Delta(3) \neq 0$ and $(\Delta(1), \Delta(2)) \rightarrow 0$. Therefore formally we have omitted the following channels of electrostatic interaction $0-1s,3-1s, 0-2s,3-2s$. It is commonly known that superconducting state especially with strong superconductivity as in case of bulk superconductor is not supporting and shielding itself from the external and internal electrostatic field of certain strength as it naturally protects its ground superconducting macroscopic state.
Having established the mathematical structure describing the electrostatic interaction between semiconductor position-based qubit and Josephson junction we can move into first analytical and numerical calculations. First simplification is that $\Delta(1)=\Delta(2)=0$ and $\Delta=\Delta(0)=\Delta(3) \in R$ so it means that there is no net electric current flowing via Josephson junction since the electric current flow imposes the condition of phase difference among superconducting order parameter $\Delta(0)$ and $\Delta(3)$ and in such case superconducting order parameter is complex valued scalar. Also it implies that there is no magnetic field in our system since magnetic field brings phase imprint between $\Delta(0)$ and $\Delta(3)$. Second simplification is that $E_{p1}=E_{p2}=E_p, t_s \in R$. Third simplification is that $E_{e0}=E_{e1}=E_{e2}=E_{e3}=-E_{h0}=-E_{h1}=-E_{h2}=-E_{h3}=V$ so it implies electron-hole symmetry in area of ABS that is the middle of Josephson junction.  In such way all hole eigenenergies are corresponding to electron eigenenergies with $-$ sign. Last assumption is that electron or hole hopping in the area of ABS in between nearest neighbours is such that $t_{e(k,k+1)} \neq 0$ and $t_{h(k,k+1)} \neq 0$ and is 0 otherwise. One can name such feature of transport in Josephson junction as diffusive and not ballistic what brings the mathematical simplifications. Having established such facts we can move into analytical and numerical calculations. The Hamiltonian of physical system has such structure that allows analytic determination of all eigenenergies since Hamiltonian matrix has many symmetries. In particular we can obtain the spectrum of eigenenergies in dependence on the distance d as depicted in Fig.3 
and spectrum of eigeneneries in dependence of superconducting order parameter as given in Fig.4. 
It is possible to observe the swap of the ground and excited state in the system of interacting Josephson junction with semiconductor qubit what implies the existence of topological phase transition. This makes such system to be interesting for quantum information processing both in classical quantum ways as with use of topological states of matter. Topological states of matter can be controlled by tunning the superconducting order parameter strenght as with use of external magnetic field or by using array of semiconductor quantum dots where the distance between single electron (distributed in wavepacket) and Josephson junctions can be changed \cite{Nbodies} with use of electric control acting on the state of semiconductor qubit.

One of the most interesting feature observed in BdGe tight-binding model is the landscape tunning of eigenenergies by application of small voltage (below the size $2 e \Delta$) to non-superconducting region of Josephson junction. In very real way the control of this voltage is the control of chemical potential in Josephson junction.

In such case one obtains various features as described in Fig. 5-7.
In the described considerations the spin degree-of-freedom was omitted in case of Josephson junction as well as in case of semiconductor position based qubit. However they could be easily included but it would increase the size of matrix describing interaction between superconductor Josephson junction and semiconductor electrostatic qubit from 16 by 16 to the size 8*4=32 so one obtains matrix 32 by 32. Adding strong spin-orbit interaction to the Hamiltonian of Josephson junction under the presence of magnetic field allows to describe topological Josephson junction. In such way we can obtain the effective 32 by 32 Hamiltonian for interaction between semiconductor position based qubit and topological Josephson junction in minimalistic way.
It shall be also underlined that so far we have used BdGe formalism that is suitable for mean field theory domain. However, in our case we have considered very special interactions between individual (electrons, holes) present in area of Josephson junction and specific individual electron present in area of semiconductor qubit. Usage of BdGe formalism is therefore first level of possible approximation and further more detailed study can be
attempted in determination of microscopic processes present interacting Josephson junction with semicondutor qubit in more detailed way. It is sufficient to mention that in our case superconductors shall have relatively small size so we are dealing with relatively small number of electrons and holes in non-superconducting area. More detailed considerations are however beyond the scope of this work and requires Density Functional Theory (DFT) methods, etc.

\begin{figure}
\centering
\label{fig:Spectrumd}
\includegraphics[scale=0.95]{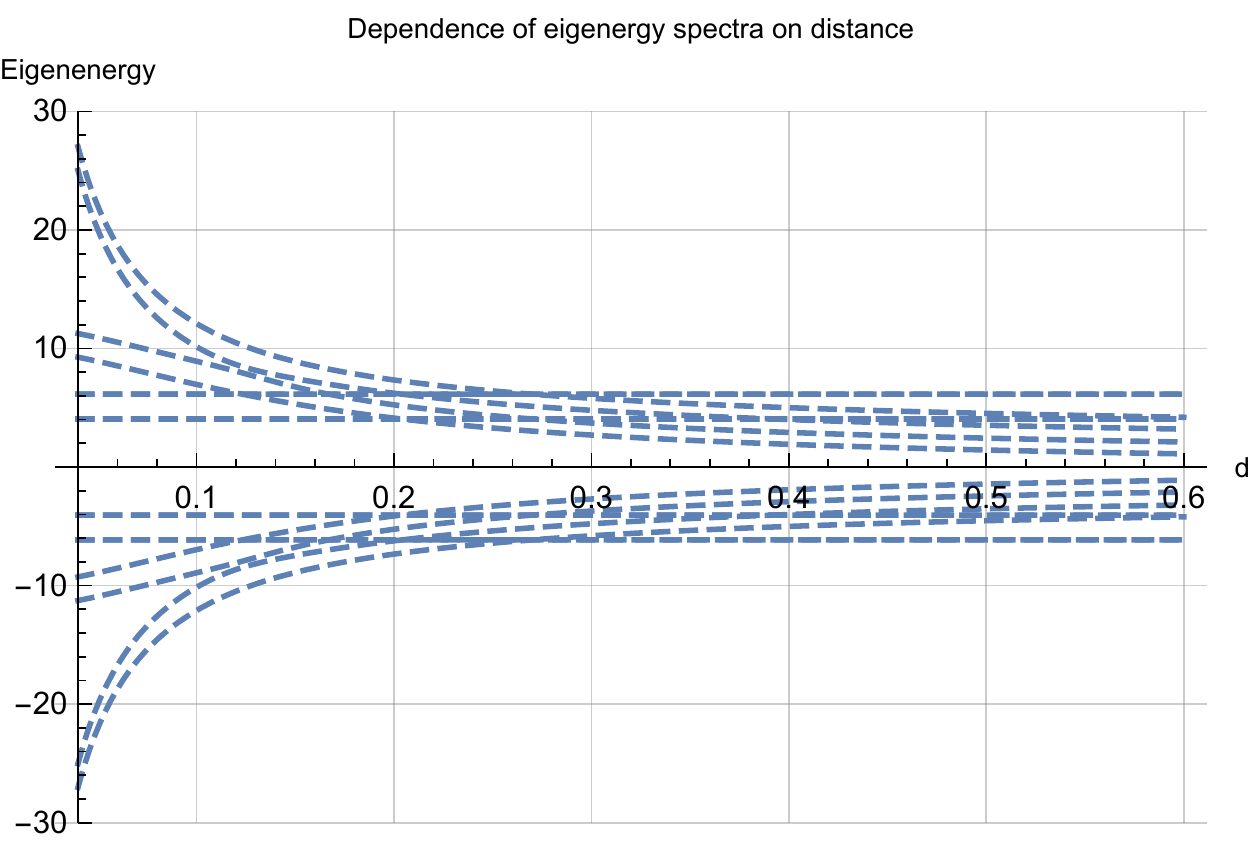} 
\caption{Eigenenergies of semiconductor qubit coupled to Josephson junction in dependence on distance in tight-binding minimalitic approach.}
\end{figure}

\begin{figure}
\centering
\label{fig:SpectrumDelta}
\includegraphics[scale=0.95]{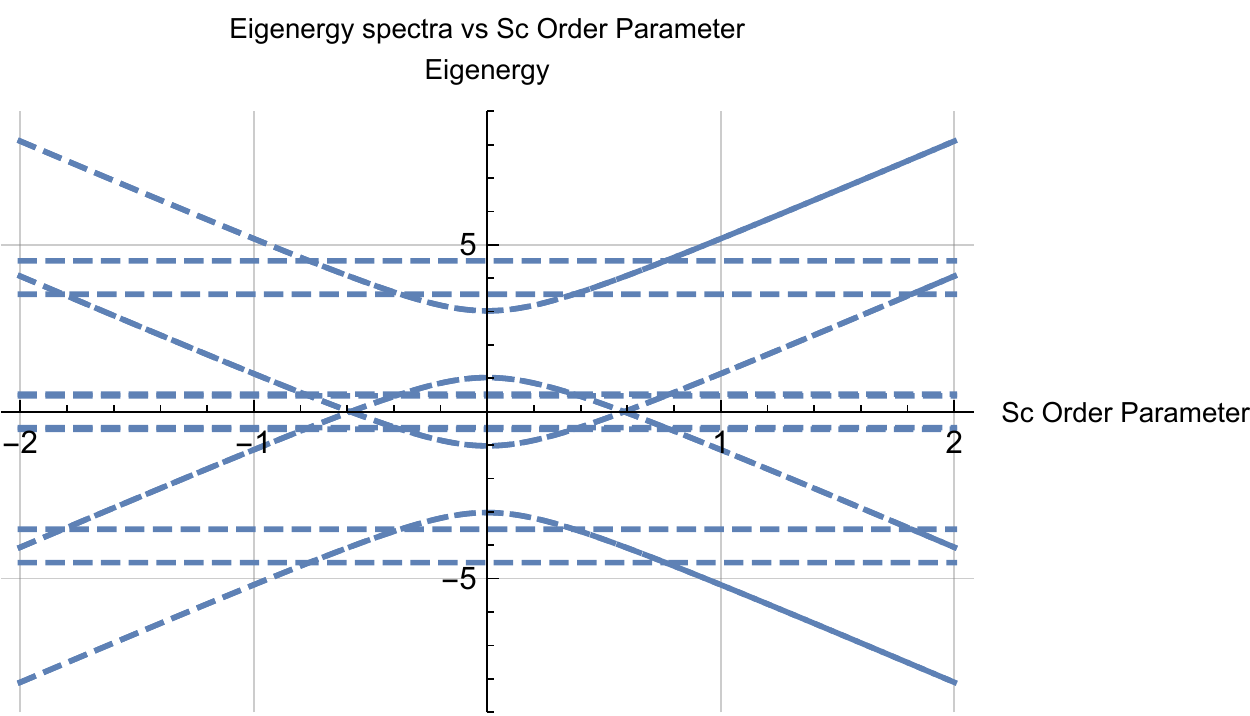} 
\caption{Eigenenergies of semiconductor qubit coupled to Josephson junction in dependence on superconducting order parameter in minimalitic approach.}
\end{figure}


\begin{figure}
\centering
\label{fig:SpectrumV}
\includegraphics[scale=1.1]{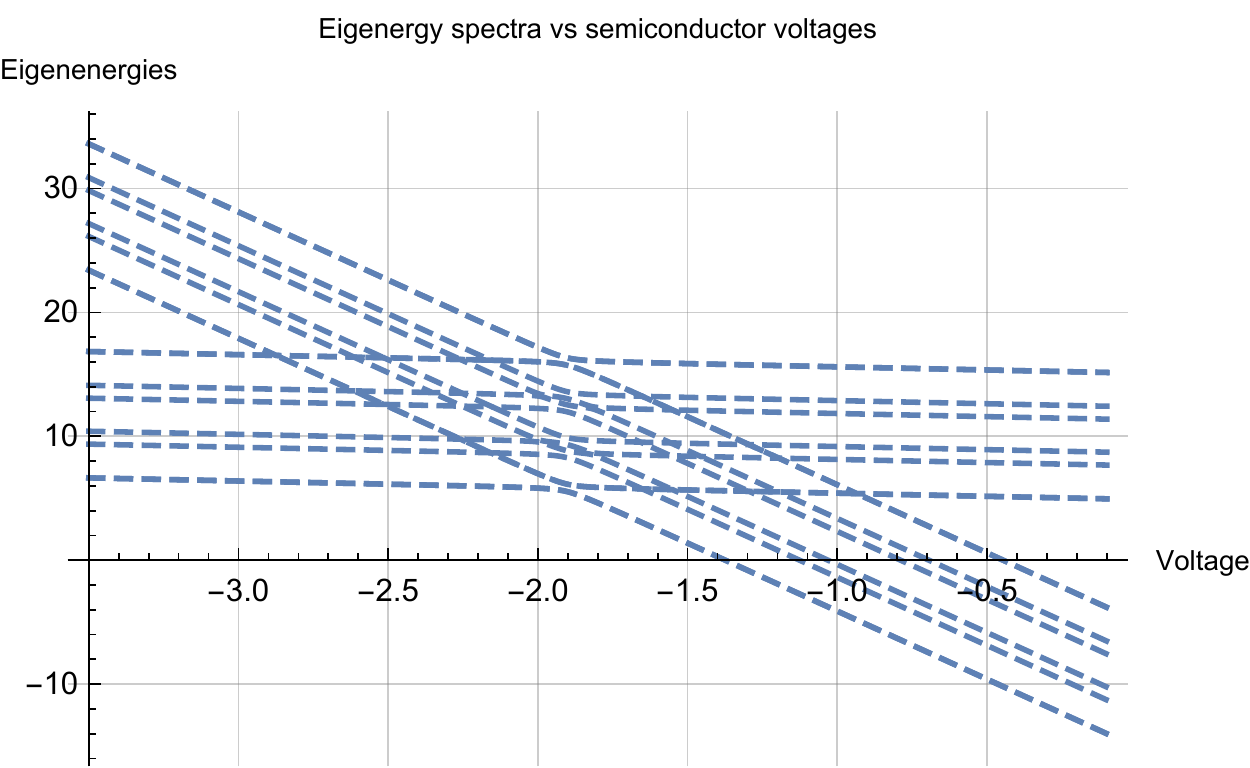} 
\caption{Tunnning the spectrum of eigenenergies in electrostatic qubit interacting with Josephson junction while we are changing the chemical potential of insulator region in Josephson junction at all nodes 0, 1, 2 and 3 in the same time.}
\centering
\label{fig:SpectrumVQ}
\includegraphics[scale=1.1]{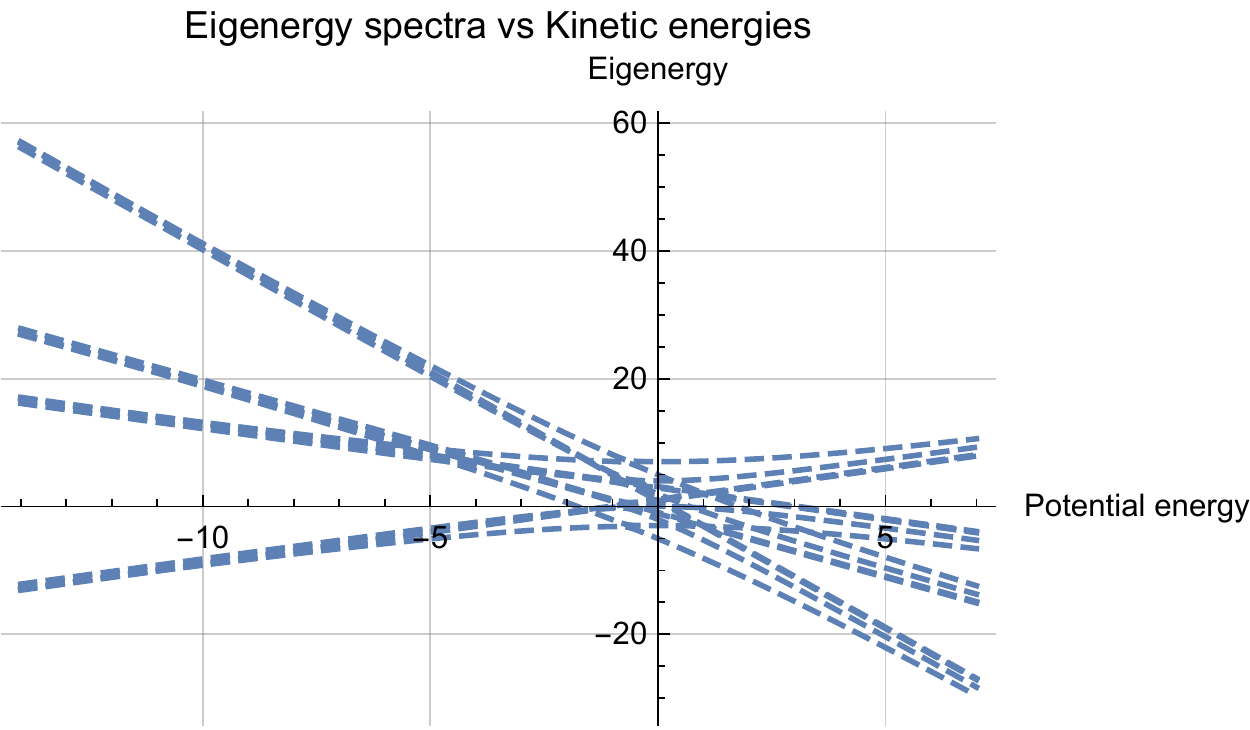}
\centering
\label{fig:SpectrumVQ1}
\includegraphics[scale=1.1]{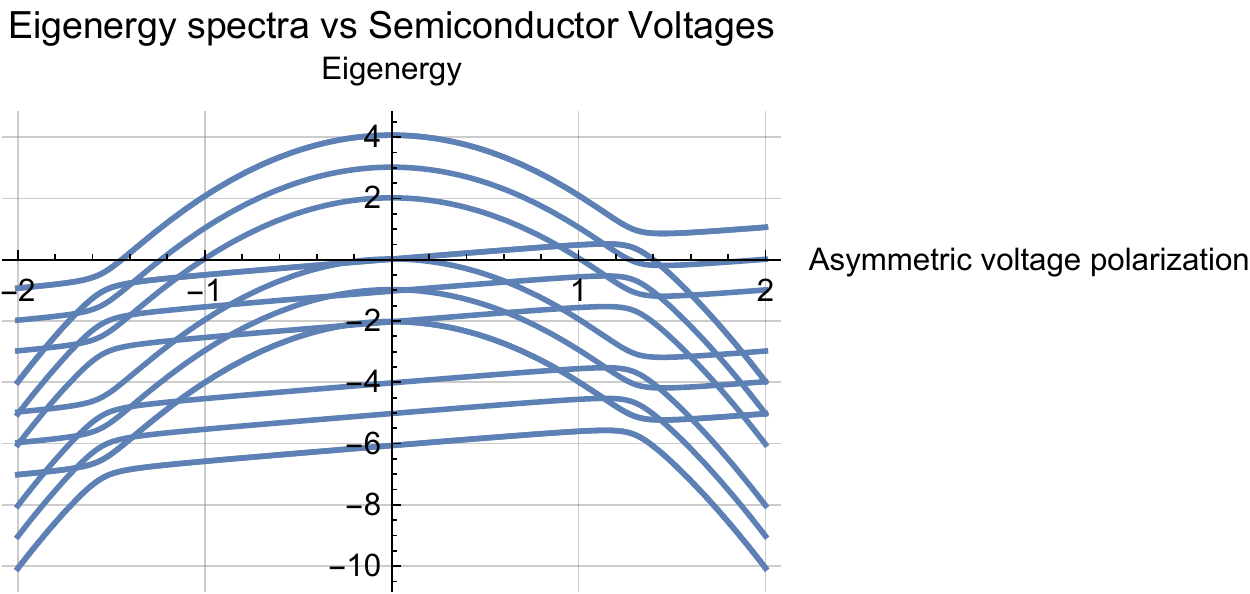}
\caption{Different eigenenergies as functions of tunning voltages. }
\end{figure}


\begin{figure}
\centering
\label{fig:SpectrumVQQ}
\includegraphics[scale=1.1]{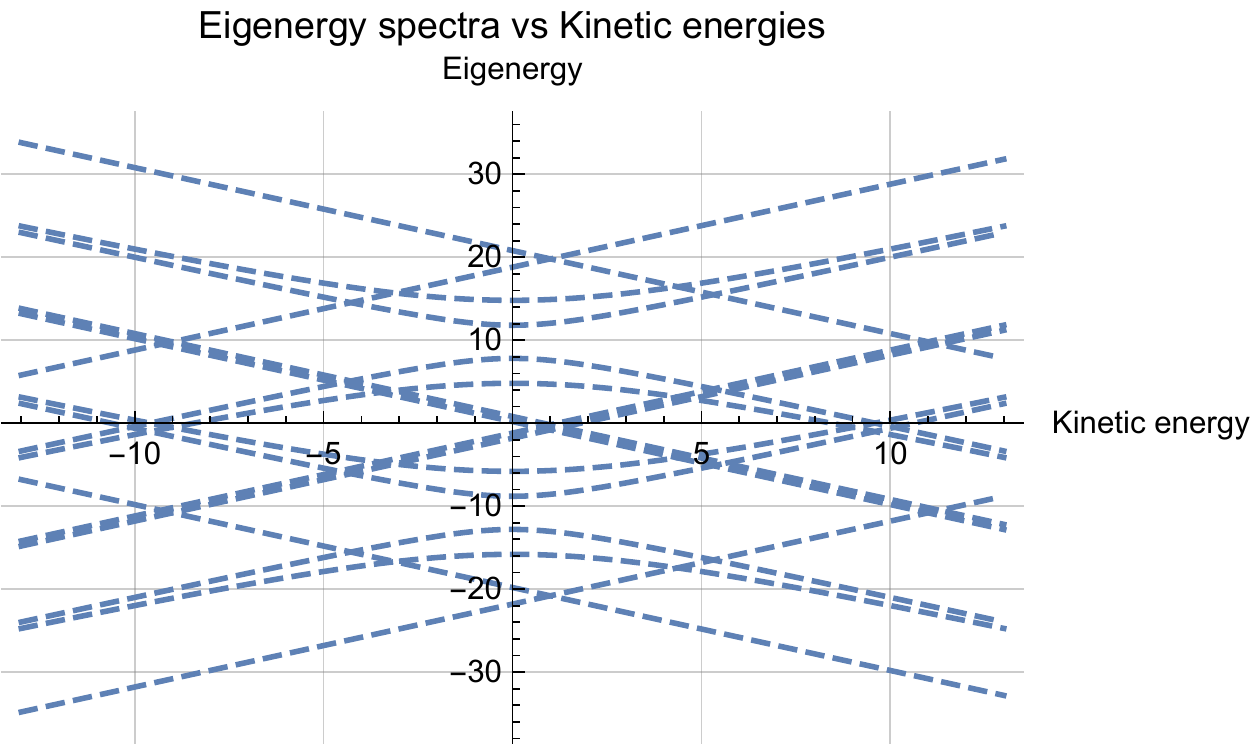}
\caption{Different eigenenergies as functions of tunning voltages. }
\end{figure}

and
\tiny
\begin{eqnarray*}
H_{eff}= 
\left(
\begin{array}{cccccccccccccccc}
 E_{p}+V & t_{s} & 0 & 0 & 0 & 0 & 0 & 0 & \Delta_1 e^{i \phi_{1}} & 0 & 0 & 0 & 0 & 0 & 0 & 0
   \\
 t_{s} & E_{p}+V & 0 & 0 & 0 & 0 & 0 & 0 & 0 & \Delta_1 e^{i \phi_{1}} & 0 & 0 & 0 & 0 & 0 & 0
   \\
 0 & 0 & E_{c1}+E_{p}+V & t_{s} & t_{j} & 0 & 0 & 0 & 0 & 0 & 0 & 0 & 0 & 0 & 0 & 0 \\
 0 & 0 & t_{s} & E_{c2}+E_{p}+V & 0 & t_{j} & 0 & 0 & 0 & 0 & 0 & 0 & 0 & 0 & 0 & 0 \\
 0 & 0 & t_{j} & 0 & E_{c2}+E_{p}+V & t_{s} & 0 & 0 & 0 & 0 & 0 & 0 & 0 & 0 & 0 & 0 \\
 0 & 0 & 0 & t_{j} & t_{s} & E_{c1}+E_{p}+V & 0 & 0 & 0 & 0 & 0 & 0 & 0 & 0 & 0 & 0 \\
 0 & 0 & 0 & 0 & 0 & 0 & E_{p}+V & t_{s} & 0 & 0 & 0 & 0 & 0 & 0 & \Delta_2 e^{i \phi_{2}} & 0
   \\
 0 & 0 & 0 & 0 & 0 & 0 & t_{s} & E_{p}+V & 0 & 0 & 0 & 0 & 0 & 0 & 0 & \Delta_2 e^{i \phi_{2}}
   \\
 \Delta_1 e^{-i \phi_{1}} & 0 & 0 & 0 & 0 & 0 & 0 & 0 & E_{p}-V & t_{s} & 0 & 0 & 0 & 0 & 0 &
   0 \\
 0 & \Delta_1 e^{-i \phi_{1}} & 0 & 0 & 0 & 0 & 0 & 0 & t_{s} & E_{p}-V & 0 & 0 & 0 & 0 & 0 &
   0 \\
 0 & 0 & 0 & 0 & 0 & 0 & 0 & 0 & 0 & 0 & - E_{c1}+E_{p}-V & t_{s} & -t_{j} & 0 & 0 & 0 \\
 0 & 0 & 0 & 0 & 0 & 0 & 0 & 0 & 0 & 0 & t_{s} & -E_{c2}+E_{p}-V & 0 & -t_{j} & 0 & 0 \\
 0 & 0 & 0 & 0 & 0 & 0 & 0 & 0 & 0 & 0 & -t_{j} & 0 & -E_{c2}+E_{p}-V & t_{s} & 0 & 0 \\
 0 & 0 & 0 & 0 & 0 & 0 & 0 & 0 & 0 & 0 & 0 & -t_{j} & t_{s} & -E_{c1}+E_{p}-V & 0 & 0 \\
 0 & 0 & 0 & 0 & 0 & 0 & \Delta_2 e^{-i \phi_{2}} & 0 & 0 & 0 & 0 & 0 & 0 & 0 & E_{p}-V &
   t_{s} \\
 0 & 0 & 0 & 0 & 0 & 0 & 0 & \Delta_2 e^{-i \phi_{2}} & 0 & 0 & 0 & 0 & 0 & 0 & t_{s} &
   E_{p}-V \\
\end{array}
\right)
\end{eqnarray*}
\normalsize
and it given eigenstates
 and eigenstates $\ket{\psi}_{1}=$ 
\tiny
     \begin{eqnarray*}
=
\begin{pmatrix}
0, \\ 0, \\ -1,\\ -\frac{2 (t_{j}-t_{s})}{\sqrt{(E_{c1}-E_{c2})^2
+4(t_{j}-t_{s})^2}-(E_{c1}-E_{c2})}, \\ -\frac{t_{j} e^{i
   \phi_{1}+i \phi_{2} } \left(t_{j}^2 e^{2 i \phi_{1}+2 i \phi_{2}}-t_{s}^2 e^{2 i
   \phi_{1}+2 i \phi_{2}}\right)-t_{j} e^{i \phi_{1}+i \phi_{2}} \left(\frac{1}{2} e^{i
   \phi_{1}+i \phi_{2}} \left(\sqrt{(E_{c1}-E_{c2})^2+4(t_{j}-t_{s})^2})-(E_{c1}+E_{c2})-2 (E_{p}+ V)\right)+e^{i \phi_{1}+i
   \phi_{2}} (E_{c1}+E_{p}+V)\right) \left(\frac{1}{2} e^{i \phi_{1}+i \phi_{2}}
   \left(\sqrt{ (E_{c1}-E_{c2})^2+4 (t_{j}^2-t_{s})^2}-(E_{c1}+E_{c2})-2 (E_{p}+ V) \right)+e^{i (\phi_{1}+\phi_{2}) }
   (E_{c2}+E_{p}+V)\right)}{-t_{j}t_{s} e^{3 i \phi_{1}+3 i \phi_{2}}
   \sqrt{(E_{c1}-E_{c2})^2+4t_{j}^2-8t_{j}t_{s}+4
   t_{s}^2}+E_{c1} t_{j}t_{s} e^{3 i \phi_{1}+3 i \phi_{2}}-E_{c2} t_{j}
   t_{s} e^{3 i \phi_{1}+3 i \phi_{2}}},\\
   1, \\ 0, \\ 0, \\ 0, \\ 0, \\ 0, \\ 0, \\ 0, \\ 0, \\ 0, \\ 0
   \end{pmatrix}, 
\end{eqnarray*}
\normalsize
\begin{eqnarray}
E_1=\frac{1}{2} \left(
   -\sqrt{ (E_{c1}^2-E_{c2})^2+4 (t_{j}- t_{s})^2}-(E_{c1}+E_{c2}) +2(E_{p}-V) \right)
\end{eqnarray}
$\ket{\psi}_{2}=$ \tiny
    \begin{eqnarray*}
    \begin{pmatrix}
0, \\
0, \\
0, \\
0, \\
0, \\
0, \\
0, \\
0, \\
0, \\
0, \\
1, \\
,-\frac{2 (t_{j}-t_{s})}{\sqrt{ (E_{c1}-E_{c2})^2+4 (t_{j}-
   t_{s})^2}-(E_{c1}-E_{c2})},\\
   -\frac{t_{j} e^{i \phi_{1}+i \phi_{2}} \left(-\frac{1}{2}
   e^{i(\phi_{1}+\phi_{2})} \left(\sqrt{(E_{c1}-E_{c2})^2+4
   (t_{j}-t_{s})^2}-(E_{c1}+E_{c2})+2(E_{p}-V)\right)-e^{i
   \phi_{1}+i\phi_{2}} (E_{c1}-E_{p}+V)\right) \left(-\frac{1}{2} e^{i \phi_{1}+i
   \phi_{2}} \left(\sqrt{(E_{c1}-E_{c2})^2+4(t_{j}-t_{s})^2}-(E_{c1}+E_{c2})+2 (E_{p}-V)\right)-e^{i(\phi_{1}+\phi_{2})}
   (E_{c2}-E_{p}+V)\right)-t_{j} e^{i(\phi_{1}+\phi_{2})} \left(t_{j}^2 e^{2 i
   \phi_{1}+2 i \phi_{2}}-t_{s}^2 e^{2 i \phi_{1}+2 i \phi_{2}}\right)}{-t_{j}t_{s}
   e^{3 i \phi_{1}+3 i \phi_{2}} \sqrt{(E_{c1}-E_{c2})^2+4
   (t_{j}- t_{s})^2}+E_{c1}t_{j}t_{s} e^{3 i \phi_{1}+3 i
   \phi_{2}}-E_{c2}t_{j}t_{s}e^{3i(\phi_{1}+\phi_{2}) }}, \\
   1, \\
   0, \\
   0
   \end{pmatrix}
\end{eqnarray*}
\normalsize
\begin{eqnarray}
E_2=\frac{1}{2}  \left(
   +\sqrt{(E_{c1}-E_{c2})^2+4(t_{j}^2-t_{s})^2}+(E_{c1}-E_{c2}) +2(E_{p}-V) \right)
\end{eqnarray}
$\ket{\psi}_{3}=$ \tiny
     \begin{eqnarray}
\begin{pmatrix}
0,\\
0, \\
1, \\
-\frac{2 (t_{j}+t_{s})}{\sqrt{(E_{c1}-E_{c2})^2+4
   (t_{j}+t_{s})^2}-(E_{c1}-E_{c2})}, \\
-\frac{t_{j} e^{i3(
   \phi_{1}+\phi_{2})} \left(t_{j}^2-t_{s}^2\right)-t_{j} e^{i \phi_{1}+i \phi_{2}} \left(\frac{1}{2} e^{i
   \phi_{1}+i \phi_{2}} \left(\sqrt{(E_{c1}-E_{c2})^2+4(t_{j}+t_{s})^2 }-(E_{c1}+E_{c2})-2 (E_{p}+V)\right)+e^{i \phi_{1}+i
   \phi_{2}} (E_{c1}+E_{p}+V)\right) \left(\frac{1}{2} e^{i \phi_{1}+i \phi_{2}}
   \left(\sqrt{(E_{c1}-E_{c2})^2+4(t_{j}+
   t_{s})^2}-(E_{c1}+E_{c2})-2(E_{p}+V)\right)+e^{i\phi_{1}+i \phi_{2}}
   (E_{c2}+E_{p}+V)\right)}{[-t_{j}t_{s} e^{3 i (\phi_{1}+\phi_{2})}
   \sqrt{(E_{c1}-E_{c2})^2+4 (t_{j}+t_{s})^2}+E_{c1}t_{j}t_{s} e^{3 i \phi_{1}+3 i\phi_{2}}-E_{c2} t_{j}
   t_{s} e^{3 i \phi_{1}+3 i \phi_{2}}]},\\1,\\ 0,\\ 0,\\ 0,\\ 0,\\ 0,\\ 0,\\ 0,\\ 0,\\ 0, \\ 0
   \end{pmatrix}
\end{eqnarray}
\normalsize
\begin{eqnarray}
E_3=\frac{1}{2} \left(-
   \sqrt{(E_{c1}-E_{c2})^2+4 (t_{j}+
    t_{s})^2}-(E_{c1}+E_{c2}) +2(E_{p}-V) \right)
\end{eqnarray}
$\ket{\psi}_{4}=$ \tiny
     \begin{eqnarray*}
\begin{pmatrix}
0, \\
0, \\
0, \\
0, \\
0, \\
0, \\
0, \\
0, \\
0, \\
0, \\
-1, \\
-\frac{2 (t_{j}+t_{s})}{\sqrt{(E_{c1}-E_{c2})^2+4(t_{j}+t_{s})^2}-E_{c1}+E_{c2}}, \\
   -\frac{t_{j} e^{i \phi_{1}+i \phi_{2}} \left(-\frac{1}{2}
   e^{i \phi_{1}+i \phi_{2}} \left(\sqrt{(E_{c1}-E_{c2})^2+4(t_{j}+t_{s})^2}-(E_{c1}+E_{c2})+2(E_{p}-V)\right)-e^{i
   \phi_{1}+i\phi_{2}} (E_{c1}-E_{p}+V)\right) \left(-\frac{1}{2} e^{i \phi_{1}+i
   \phi_{2}} \left(\sqrt{(E_{c1}-E_{c2})^2+4 (t_{j}+t_{s})^2}-(E_{c1}+E_{c2})+2 (E_{p}-V)\right)-e^{i \phi_{1}+i \phi_{2}}
   (E_{c2}-E_{p}+V)\right)-t_{j} e^{i 3(\phi_{1}+\phi_{2})} \left(t_{j}^2 -t_{s}^2 \right)}{-t_{j}t_{s}
   e^{3 i (\phi_{1}+\phi_{2})} \sqrt{ (E_{c1}-E_{c2})^2+4
   (t_{j}+t_{s})^2}+E_{c1} t_{j}t_{s} e^{3 i \phi_{1}+3 i
   \phi_{2}}-E_{c2} t_{j}t_{s} e^{3 i \phi_{1}+3 i \phi_{2}}}, \\
   1, \\
   0, \\
   0
   \end{pmatrix}
\end{eqnarray*}

\normalsize
\begin{eqnarray}
E_4=
\frac{1}{2}\left(
   \sqrt{+(E_{c1}-E_{c2})^2+4(t_{j}+t_{j})}-(E_{c1}+E_{c2}) +2 (E_{p} - V) \right)
\end{eqnarray}
$ \ket{\psi}_{5}= $ \tiny
     \begin{eqnarray*}
\begin{pmatrix}
0, \\
0,\\
0,\\
0,\\
0,\\
0,\\
0,\\
0,\\
0,\\
0,\\
1, \\
\frac{2 (t_{j}-t_{s})}{\sqrt{ (E_{c1}-E_{c2})^2+4 (t_{j}-t_{s})^2}+(E_{c1}-E_{c2})},\\
   -\frac{t_{j} e^{i \phi_{1}+i \phi_{2}} \left(\frac{1}{2}
   e^{i \phi_{1}+i \phi_{2}} \left(\sqrt{(E_{c1}-E_{c2})^2+4
   (t_{j}-t_{s})^2}+(E_{c1}+E_{c2})-2(E_{p}- V)\right)-e^{i
   \phi_{1}+i \phi_{2}} (E_{c1}-E_{p}+V)\right) \left(\frac{1}{2} e^{i \phi_{1}+i
   \phi_{2}} \left(\sqrt{(E_{c1}-E_{c2})^2+4 (t_{j}-t_{s})^2}+(E_{c1}+E_{c2})-2 (E_{p}- V)\right)-e^{i (\phi_{1}+\phi_{2})}
   (E_{c2}-E_{p}+V)\right)-t_{j} \left(t_{j}^2 e^{3 i(
   \phi_{1}+\phi_{2})}-t_{s}^2 e^{3 i (\phi_{1}+\phi_{2})}\right)}{t_{j}t_{s}
   e^{3 i \phi_{1}+3 i \phi_{2}} \sqrt{(E_{c1}-E_{c2})^2+4
   (t_{j}-t_{s})^2}+E_{c1} t_{j}t_{s} e^{3 i (\phi_{1}+
   \phi_{2})}-E_{c2} t_{j} t_{s} e^{3 i \phi_{1}+3 i \phi_{2}}},\\
   1, \\
   0, \\
   0
   \end{pmatrix},
\end{eqnarray*}
\normalsize
\begin{eqnarray}
E_5=
\frac{1}{2}  \left(-
   \sqrt{(E_{c1}^2-E_{c2})^2+4 (t_{j}-t_{s})^2}+(E_{c1} +E_{c2}) +2
   (E_{p} +V) \right),
\end{eqnarray}
$\ket{\psi}_{6}=$
\tiny
     \begin{eqnarray*}
\begin{pmatrix}
0, \\
0, \\
-1, \\
\frac{2 (t_{j}-t_{s})}{\sqrt{ (E_{c1}-E_{c2})^2+4
   (t_{j}-t_{s})^2}+(E_{c1}-E_{c2})}, \\
   -\frac{t_{j} e^{i
   \phi_{1}+i \phi_{2}} \left(t_{j}^2 e^{2 i \phi_{1}+2 i \phi_{2}}-t_{s}^2 e^{2 i
   \phi_{1}+2 i \phi_{2}}\right)-t_{j} e^{i \phi_{1}+i \phi_{2}} \left(e^{i \phi_{1}+i
   \phi_{2}} (E_{c1}+E_{p}+V)-\frac{1}{2} e^{i \phi_{1}+i \phi_{2}}
   \left(\sqrt{(E_{c1}-E_{c2})^2+4(t_{j}-t_{s})^2}+(E_{c1}+E_{c2})+2(E_{p}+V)\right)\right) \left(e^{i \phi_{1}+i \phi_{2}}
   (E_{c2}+E_{p}+V)-\frac{1}{2} e^{i \phi_{1}+i \phi_{2}} \left(\sqrt{(E_{c1}-E_{c2})^2+4 (t_{j}-t_{s})^2}+(E_{c1}+E_{c2})+2
   E_{p}+2 V\right)\right)}{t_{j}t_{s} e^{3 i \phi_{1}+3 i \phi_{2}} \sqrt{(E_{c1}-E_{c2})^2+4(t_{j}-t_{s})^2}+E_{c1} t_{j}
   t_{s} e^{3 i \phi_{1}+3 i \phi_{2}}-E_{c2} t_{j} t_{s} e^{3 i \phi_{1}+3 i
   \phi_{2}}},\\ 1,\\ 0,\\ 0,\\ 0,\\ 0,\\ 0,\\ 0,\\ 0, \\ 0,\\ 0, \\0
\end{pmatrix},
\end{eqnarray*}
\normalsize
\begin{eqnarray}
E_6=
\frac{1}{2} \left( \sqrt{(E_{c1}-E_{c2})^2+4(t_{j}-t_{s})^2}+(E_{c1}+E_{c2})+2
   (E_{p}+V)\right), \nonumber \\ E_7=
\frac{1}{2}\left(-\sqrt{(E_{c1}-E_{c2})^2+4 (t_{j}+
   t_{s})^2}+(E_{c1} +E_{c2})+2
   E_{p} +2 V \right), 
\end{eqnarray}
$\ket{\psi}_{7}= $
\tiny
     \begin{eqnarray*}
\begin{pmatrix}
0,\\ 0,\\ 0,\\ 0,\\ 0,\\ 0,\\ 0,\\ 0,\\ 0,\\ 0,\\ -1,\\ \frac{2 (t_{j}+t_{s})}{\sqrt{ (E_{c1}-E_{c2})^2+4 (t_{j}+
   t_{s})^2}+E_{c1}-E_{c2}},\\ -\frac{t_{j} e^{i \phi_{1}+i \phi_{2}} \left(\frac{1}{2}
   e^{i \phi_{1}+i \phi_{2}} \left(\sqrt{(E_{c1}-E_{c2})^2+4
   (t_{j}+t_{s})^2}+(E_{c1}+E_{c2})-2(E_p-V)\right)-e^{i
   \phi_{1}+i \phi_{2}} (E_{c1}-E_{p}+V)\right) \left(\frac{1}{2} e^{i \phi_{1}+i
   \phi_{2}} \left(\sqrt{(E_{c1}-E_{c2})^2+4(t_{j}+t_{s})^2}+E_{c1}+E_{c2}-2(E_p-V)\right)-e^{i \phi_{1}+i \phi_{2}}
   (E_{c2}-(E_{p}-V))\right)-t_{j} e^{i \phi_{1}+i \phi_{2}} \left(t_{j}^2 e^{2 i
   \phi_{1}+2 i \phi_{2}}-t_{s}^2 e^{2 i \phi_{1}+2 i \phi_{2}}\right)}{t_{j}t_{s}
   e^{3 i \phi_{1}+3 i \phi_{2}} \sqrt{(E_{c1}-E_{c2})^2+4
   (t_{j}+t_{s})^2 }+E_{c1} t_{j}t_{s} e^{3 i (\phi_{1}+
   \phi_{2})}-E_{c2} t_{j}t_{s} e^{3 i \phi_{1}+3 i \phi_{2}}},\\ 1,\\ 0,\\ 0
   \end{pmatrix}, \nonumber \\
\end{eqnarray*}

\normalsize
 $  E_8=\frac{1}{2} \left(+\sqrt{(E_{c1}-E_{c2})^2+4 (t_{j}+t_{j})^2}+(E_{c1}+E_{c2})
 +2 (E_{p} +V)\right)$
\small
     \begin{eqnarray*}
\ket{\psi}_{8}= \nonumber \\ \tiny
\begin{pmatrix}
0, \\
0, \\
1, \\
\frac{2 (t_{j}+t_{s})}{\sqrt{(E_{c1}-E_{c2})^2+4
   (t_{j}+t_{s})^2}+(E_{c1}-E_{c2})}, \\
   -\frac{t_{j} e^{i
   (\phi_{1}+\phi_{2})} \left(t_{j}^2 e^{2i(\phi_{1}+\phi_{2})}-t_{s}^2 e^{2 i
   (\phi_{1}+\phi_{2}) }\right)-t_{j}\left(e^{i 2\phi_{1}+i 2
   \phi_{2}} (E_{c1}+E_{p}+V)-\frac{1}{2} e^{i 2\phi_{1}+i 2\phi_{2}}
   \left(\sqrt{(E_{c1}-E_{c2})^2+4 (t_{j}+t_{s})^2}+(E_{c1}+E_{c2})+2E_{p}+2 V\right)\right) \left(e^{i \phi_{1}+i \phi_{2}}
   (E_{c2}+E_{p}+V)-\frac{1}{2} e^{i \phi_{1}+i \phi_{2}} \left(\sqrt{(E_{c1}-E_{c2})^2+4 (t_{j}+t_{s})^2 }+(E_{c1}+E_{c2})+2
   E_{p}+2 V\right)\right)}{t_{j}t_{s} e^{3 i \phi_{1}+3 i \phi_{2}} \sqrt{ (E_{c1}-E_{c2})^2+4 (t_{j}+t_{s})^2}+E_{c1} t_{j}
   t_{s} e^{3 i(\phi_{1}+\phi_{2})}-E_{c2} t_{j} t_{s} e^{3 i (\phi_{1}+\phi_{2})}}, \\
   1, \\
   0, \\
   0, \\
   0, \\
   0, \\
   0, \\
   0, \\
   0, \\
   0, \\
   0, \\
   0 \\
   \end{pmatrix}, 
 \end{eqnarray*}
 \normalsize
 We recognize that state states $\ket{\psi_1}, .. , \ket{\psi_8}$ are entangled due to non-zero Coulomb interaction between semiconductor position-based qubit and Josephson junction superconducting qubit. However other states are not entangled and are given as tensor product of two non-interacting quantum systems (what is equivalent to semiconductor qubit and superconducting qubit at sufficiently high distances) and what is also reflected in the lack of dependence of eigenenergies on Coulomb energy. Quick evaluation of energies involved in BdGe tight-binding model are specified in Table I. For the sake of comparison the length of Josephson junction was assumed to be 100nm (smaller than superconducting coherence length for most low temperature BCS superconductors) as well as size of position based qubit was assumed to be 100 nm as well (most recent technologies allows for reduction of this size to 3nm).
 \begin{figure}
 \begin{tabular}{|l|l|l|}
\hline
\textbf{BdGe tight binding parameter} &\textbf{ Mathematical formula} & \textbf{Physical value} \emph{($\Delta x=d=100nm$, n=m=1)} \\ \hline
$t_j$ & $\frac{\hbar^2}{2m_{e,sc}} (2n+1) (\frac{2\pi}{d_{semi-qbit}})^2=\frac{\hbar^2}{2m_{e,sc}} (2n+1) (\frac{2\pi}{\Delta x})^2$ & =3*5.938meV=17.814 meV \\ \hline
$t_s$ & $\frac{\hbar^2}{2m_{e,semi}} (2k+1) (\frac{2\pi}{d_{JJ}})^2 = \frac{\hbar^2}{2m_{e,semi}} (2k+1) (\frac{2\pi}{\Delta x})^2$  & =3*5.938 meV=17.814 meV  \\ \hline
$E_{c1}$ & $\frac{q^2}{a}=\frac{q^2}{4 \pi \epsilon_0 d}=\frac{e^2}{4 \pi \epsilon_0 \Delta x}$  & =0.145meV \\ \hline
$E_{c2}$ & $\frac{q^2}{b}=\frac{q^2}{4 \pi \epsilon_0 \sqrt{d^2+(2 \Delta x)^2}}=\frac{e^2}{4 \sqrt{5} \pi \epsilon_0 \Delta x} $  & =0.0659 meV \\
\hline
\end{tabular}
Table I: Scaling of tight-binding model parameters with geometry of interface between semiconductor quantum dot qubit and Josephson junction qubit in symmetric case. 
\end{figure}
 From brief analysis conducted in Table 1 one can conclude that it is desired to use long Josephson junctions that are in close proximity to semiconductor qubit so energy of kinetic excitations can be as small as possible and hence Coulomb interaction will become stronger tunning factor. Using strong superconductors would compensate the electrostatic qubit-qubit interaction and thus is not desirable so one shall stay with BCS superconductors. Operational temperature shall be kept in mK regime.

  We continue the analysis of energy eigenstates of semiconductor qubit - Josephson junction system so we obtain the following non-entangled states $\ket{\psi_9}, .. , \ket{\psi_{16}}$ given explicitly as
  \begin{eqnarray}
\ket{\psi}_{9}=
\begin{pmatrix}
0, \\
0, \\
0, \\
0, \\
0, \\
0, \\
-\frac{e^{-i \phi_{1}} \left(V e^{i \phi_{1}+i
   \phi_{2}}-\sqrt{\left(\Delta^2+V^2\right) e^{2 i \phi_{1}+2 i
   \phi_{2}}}\right)}{\Delta}, \\
   \frac{e^{-i \phi_{1}} \left(V e^{i \phi_{1}+i
   \phi_{2}}-\sqrt{\left(\Delta^2+V^2\right) e^{2 i \phi_{1}+2 i
   \phi_{2}}}\right)}{\Delta}, \\
   0,\\ 0,\\ 0,\\ 0,\\ 0,\\ 0,\\ -1,\\ 1
\end{pmatrix}, 
\ket{\psi}_{10}=
\begin{pmatrix}
-\frac{e^{-i \phi_{2}} \left(V e^{i \phi_{1}+i
   \phi_{2}}-\sqrt{\left(\Delta^2+V^2\right) e^{2 i \phi_{1}+2 i
   \phi_{2}}}\right)}{\Delta}, \\
   \frac{e^{-i \phi_{2}} \left(V e^{i \phi_{1}+i
   \phi_{2}}-\sqrt{\left(\Delta^2+V^2\right) e^{2 i \phi_{1}+2 i
   \phi_{2}}}\right)}{\Delta},\\0,\\0,\\0,\\0,\\0,\\0,\\-1,\\ 1,\\ 0,\\ 0,\\ 0,\\ 0,\\ 0,\\ 0
\end{pmatrix}, \nonumber \\
E_9=E_{10}=
\left(-\sqrt{\Delta^2 +V^2 }+E_{p} -t_{s} \right), \nonumber \\
\end{eqnarray}
\small
  \begin{eqnarray}
\ket{\psi}_{11}=
\begin{pmatrix}
0, \\
0, \\
0,\\
0, \\
0, \\
0,\\
\frac{e^{-i \phi_{1}} \left(V e^{i \phi_{1}+i
   \phi_{2}}-\sqrt{\left(\Delta^2+V^2\right) e^{2 i \phi_{1}+2 i
   \phi_{2}}}\right)}{\Delta}, \\
\frac{e^{-i \phi_{1}} \left(V e^{i \phi_{1}+i
   \phi_{2}}-\sqrt{\left(\Delta^2+V^2\right) e^{2 i \phi_{1}+2 i
   \phi_{2}}}\right)}{\Delta},\\
   0, \\
   0,\\
   0,\\
   0,\\
   0,\\
   0, \\
   1, \\
   1
   \end{pmatrix}, 
\ket{\psi}_{12}=
\begin{pmatrix}
\frac{e^{-i \phi_{2}} \left(V e^{i \phi_{1}+i \phi_{2}}-\sqrt{\left(\Delta^2+V^2\right)
   e^{2 i \phi_{1}+2 i \phi_{2}}}\right)}{\Delta}, \\
   \frac{e^{-i \phi_{2}} \left(V e^{i
   \phi_{1}+i \phi_{2}}-\sqrt{\left(\Delta^2+V^2\right) e^{2 i \phi_{1}+2 i
   \phi_{2}}}\right)}{\Delta},\\ 0,\\ 0,\\ 0,\\ 0,\\ 0,\\ 0, \\ 1,\\ 1,\\ 0,\\ 0,\\ 0,\\ 0, \\ 0, \\ 0 \\
\end{pmatrix}, \nonumber \\
E_{11}=E_{12}= \left(-\sqrt{\Delta^2 +V^2 }+E_{p} +t_{s} \right), \nonumber \\
\ket{\psi}_{13}=
\begin{pmatrix}
0, \\
0, \\
0, \\
0, \\
0, \\
0, \\
-\frac{e^{-i \phi_{1}} \left(\sqrt{\left(\Delta^2+V^2\right) e^{2 i \phi_{1}+2
   i \phi_{2}}}+V e^{i \phi_{1}+i \phi_{2}}\right)}{\Delta}, \\
   \frac{e^{-i \phi_{1}}
   \left(\sqrt{\left(\Delta^2+V^2\right) e^{2 i \phi_{1}+2 i \phi_{2}}}+V e^{i \phi_{1}+i
   \phi_{2}}\right)}{\Delta},\\ 0, \\ 0, \\ 0, \\ 0, \\  0, \\ 0, \\ -1, \\ 1
\end{pmatrix},
\ket{\psi}_{14}=
\begin{pmatrix}
-\frac{e^{-i \phi_{2}} \left(\sqrt{\left(\Delta^2+V^2\right) e^{2 i \phi_{1}+2 i
   \phi_{2}}}+V e^{i \phi_{1}+i \phi_{2}}\right)}{\Delta} \\ ,\frac{e^{-i \phi_{2}}
   \left(\sqrt{\left(\Delta^2+V^2\right) e^{2 i \phi_{1}+2 i \phi_{2}}}+V e^{i \phi_{1}+i
   \phi_{2}}\right)}{\Delta},\\ 0,\\ 0,\\ 0,\\ 0,\\ 0,\\ 0,\\ -1, \\ 1 \\,0 \\ ,0,\\ 0,\\ 0,\\ 0, \\ 0
   \end{pmatrix}, \nonumber \\
  E_{13}=E_{14}=
 \left(\sqrt{\Delta^2 +V^2 }+E_{p} -t_{s} \right)
\end{eqnarray}
\small
  \begin{eqnarray*}
\ket{\psi}_{15}=
\begin{pmatrix}
0 \\ 0 \\ 0 \\ 0 \\ 0 \\ 0 \\ \frac{e^{-i \phi_{1}} \left(\sqrt{\left(\Delta^2+V^2\right) e^{2 i \phi_{1}+2
   i \phi_{2}}}+V e^{i \phi_{1}+i \phi_{2}}\right)}{\Delta}\\ \frac{e^{-i \phi_{1}}
   \left(\sqrt{\left(\Delta^2+V^2\right) e^{2 i \phi_{1}+2 i \phi_{2}}}+V e^{i \phi_{1}+i
   \phi_{2}}\right)}{\Delta} \\ 0 \\ 0 \\ 0 \\ 0 \\ 0 \\ 0 \\ 1 \\ 1
   \end{pmatrix}, 
\ket{\psi}_{16}=
\begin{pmatrix}
\frac{e^{-i \phi_{2}} \left(\sqrt{\left(\Delta^2+V^2\right) e^{2 i \phi_{1}+2 i
   \phi_{2}}}+V e^{i \phi_{1}+i \phi_{2}}\right)}{\Delta} \\ \frac{e^{-i \phi_{2}}
   \left(\sqrt{\left(\Delta^2+V^2\right) e^{2 i \phi_{1}+2 i \phi_{2}}}+V e^{i \phi_{1}+i
   \phi_{2}}\right)}{\Delta}\\ 0 \\0 \\ 0 \\ 0 \\ 0 \\ 0 \\ 1 \\ 1 \\ 0 \\ 0 \\ 0 \\ 0 \\ 0 \\ 0
\end{pmatrix}, E_{15}=E_{16}=
 \left(\sqrt{\Delta^2 +V^2 }+E_{p} +t_{s}\right)
\end{eqnarray*}
\normalsize
\section{Conclusions}
The presented work describes elementary but still meaningful model of electrostatic interface between electrostatic position based qubit implemented in
coupled semiconductor quantum dots (as present in CMOS technology) coupled to Josephson junction qubit. The emergence of electrostatic entanglement was shown what is the example of interface between superconducting quantum computer and semiconductor quantum computer.
The obtained results have its meaning in the development of single-electron electrostatic quantum neural networks, quantum gates, such as CNOT, SWAP, Toffoli and Fredkin gates as well as any other types of quantum gates with $N$ inputs and $M$ outputs. Single-electron semiconductor devices can be attractive from point of view of power consumption and they can approach similar performance as Rapid Single Quantum Flux superconducting circuits \cite{Pomorski_spie} having much smaller dimensions than superconducting circuits. In conducted computations the spin degree-of-freedom was neglected. However it can be added in straightforward way doubling the size of Hilbert space.  The obtained results allow us to obtain the entanglement of qubit A (for example) using biparticle Von Neumann entropy $S(t)_A$ of qubit A in two electrostatically interacting qubits with time as given by formula
\begin{equation}
\label{entropyS}
S(t)=-Tr[\hat{\rho_A(t)}(\log(\hat{\rho_A}(t)))],
\end{equation}
where Tr[.] is matrix trace operator and $\rho_A$ is the reduced density matrix of A qubit after presence of B qubit was traced out. The obtained results can be mapped to Schr\"odinger formalism \cite{Xu} in order to obtain higher accuracy and resolution in description of quantum state dynamics. One can use the obtained results in determination of quantum transport in the single electron devices or arbitrary topology, which can be helpful in optimization of device functionality and sequence of controlling sequences shaping the electron confinement potential. Topological phase transitions as described by \cite{QPT}, \cite{Choi}, \cite{Belzig} are expected to take place in arrays of coupled electrostatic qubits due to the similarity of tight-binding applied in semiconductor coupled quantum well model to Josephson model in Cooper pair box superconducting qubits. All results are quite straightforward to be generalized for electrons and holes confined in net of coupled quantum dots (which changes only sign of electrostatic energy so $q^2 \rightarrow -q^2$) under the assumption that recombination processes do not occur. What is more the interaction between electrostatic position based qubit and Josephson junction was formulated and solved in tight-binding model. In quite straightforward way one obtains the electrostatically coupled networks of graphs interacting with single Josephson junction in analytical way.
It shall have its importance in the development of interface between semiconductor CMOS quantum computer and already developed superconducting computer.

\section{Acknowledgment}

This work was supported by Science Foundation Ireland under Grant 14/RP/I2921. We would like to thank to  Erik Staszewski $(erik.staszewski@ucd.ie)$ for his assistance in graphical design of figures. 

\end{document}